\documentstyle[psfig]{mn}

\newif\ifAMStwofonts
\AMStwofontstrue

\def\be{\begin{equation}}
\def\ee{\end{equation}}

\def\etal{{\it et al.~}}

\def\gs{\mathrel{\raise1.16pt\hbox{$>$}\kern-7.0pt
\lower3.06pt\hbox{{$\scriptstyle \sim$}}}}
\def\ls{\mathrel{\raise1.16pt\hbox{$<$}\kern-7.0pt
\lower3.06pt\hbox{{$\scriptstyle \sim$}}}}
\def\gtsima{$\; \buildrel > \over \sim \;$}
\def\ltsima{$\; \buildrel < \over \sim \;$}
\def\prosima{$\; \buildrel \propto \over \sim \;$}
\def\gsim{\lower.5ex\hbox{\gtsima}}
\def\lsim{\lower.5ex\hbox{\ltsima}}
\def\simgt{\lower.5ex\hbox{\gtsima}}
\def\simlt{\lower.5ex\hbox{\ltsima}}
\def\simpr{\lower.5ex\hbox{\prosima}}

\def\ga{\gsim}

\def\pp{\noindent\parshape 2 0truecm 17truecm 2truecm 15truecm}
\def\rf#1;#2;#3;#4 {\par\pp#1, #2, #3, #4. \par}

\def\pr{\ref@jnl{Phys.Rev}}

\def\ie{{\frenchspacing\it i.e. }}
\def\eg{{\frenchspacing\it e.g. }}

\def\href#1;#2 {{\bf #1} : {\em #2}}

\def\beq#1{\begin{equation}\label{#1}}
\def\eeq{\end{equation}}
\def\beqa#1{\begin{eqnarray}\label{#1}}
\def\eeqa{\end{eqnarray}}

\def\H2p{H$_2^+$ }

\def\mH2p{H_2^+}

\ifoldfss
  \ifCUPmtlplainloaded \else
    \NewTextAlphabet{textbfit} {cmbxti10} {}
    \NewTextAlphabet{textbfss} {cmssbx10} {}
    \NewMathAlphabet{mathbfit} {cmbxti10} {} 
    \NewMathAlphabet{mathbfss} {cmssbx10} {} 
  \fi
  \ifAMStwofonts
    \ifCUPmtlplainloaded \else
      \NewSymbolFont{upmath} {eurm10}
      \NewSymbolFont{AMSa} {msam10}
      \NewMathSymbol{\upi}     {0}{upmath}{19}
      \NewMathSymbol{\umu}     {0}{upmath}{16}
      \NewMathSymbol{\upartial}{0}{upmath}{40}
      \NewMathSymbol{\leqslant}{3}{AMSa}{36}
      \NewMathSymbol{\geqslant}{3}{AMSa}{3E}

    \fi
  \fi
\fi 

\ifnfssone
  \newmathalphabet{\mathit}
  \addtoversion{normal}{\mathit}{cmr}{m}{it}
  \addtoversion{bold}{\mathit}{cmr}{bx}{it}
  \newmathalphabet{\mathbfit} 
  \addtoversion{normal}{\mathbfit}{cmr}{bx}{it}
  \addtoversion{bold}{\mathbfit}{cmr}{bx}{it}
  \newmathalphabet{\mathbfss} 
  \addtoversion{normal}{\mathbfss}{cmss}{bx}{n}
  \addtoversion{bold}{\mathbfss}{cmss}{bx}{n}
  \ifAMStwofonts
    \ifCUPmtlplainloaded \else
      \UseAMStwoboldmath
      \makeatletter
      \new@mathgroup\upmath@group
      \define@mathgroup\mv@normal\upmath@group{eur}{m}{n}
      \define@mathgroup\mv@bold\upmath@group{eur}{b}{n}
      \edef\UPM{\hexnumber\upmath@group}
      \new@mathgroup\amsa@group
      \define@mathgroup\mv@normal\amsa@group{msa}{m}{n}
      \define@mathgroup\mv@bold\amsa@group{msa}{m}{n}
      \edef\AMSa{\hexnumber\amsa@group}
      \makeatother
      \mathchardef\upi="0\UPM19
      \mathchardef\umu="0\UPM16
      \mathchardef\upartial="0\UPM40
      \mathchardef\leqslant="3\AMSa36
      \mathchardef\geqslant="3\AMSa3E
    \fi
  \fi
\fi 

\ifnfsstwo
  \DeclareMathAlphabet{\mathbfit}{OT1}{cmr}{bx}{it}
  \SetMathAlphabet\mathbfit{bold}{OT1}{cmr}{bx}{it}
  \DeclareMathAlphabet{\mathbfss}{OT1}{cmss}{bx}{n}
  \SetMathAlphabet\mathbfss{bold}{OT1}{cmss}{bx}{n}
  \ifAMStwofonts
    \ifCUPmtlplainloaded \else
      \DeclareSymbolFont{UPM}{U}{eur}{m}{n}
      \SetSymbolFont{UPM}{bold}{U}{eur}{b}{n}
      \DeclareSymbolFont{AMSa}{U}{msa}{m}{n}
      \DeclareMathSymbol{\upi}{0}{UPM}{"19}
      \DeclareMathSymbol{\umu}{0}{UPM}{"16}
      \DeclareMathSymbol{\upartial}{0}{UPM}{"40}
      \DeclareMathSymbol{\leqslant}{3}{AMSa}{"36}
      \DeclareMathSymbol{\geqslant}{3}{AMSa}{"3E}
    \fi
  \fi
\fi 

\ifCUPmtlplainloaded \else
  \ifAMStwofonts \else 
    \def\upi{\pi}
    \def\umu{\mu}
    \def\upartial{\partial}
  \fi
\fi

\title[Intergalactic Scintillation]{Scintillation as a
Probe of the Intergalactic Medium}

\author[A. Ferrara \& R. Perna]{Andrea Ferrara$^{1,2}$ and Rosalba Perna$^{3,4}$ \\
$^1$ Osservatorio Astrofisico di Arcetri, Largo Enrico Fermi 5,
      50125 Firenze, Italy \\
$^2$ Center for Computational Physics, University of Tsukuba, Tsukuba-shi, Ibara
ki-ken, 305-8577, Japan\\
$^3$ Center for Astrophysics, 60 Garden Street, Cambridge, MA 02138, USA\\
$^4$ Harvard Society of Fellows}

\date{August 2000}

\pagerange{\pageref{firstpage}--\pageref{lastpage}}
\pubyear{2000}
\begin{document}

\maketitle
\label{firstpage}
\begin{abstract}
Most of the baryons in the low-redshift universe reside in a warm/hot
component which is difficult to detect with standard
absorption/emission line techniques.  We propose to use quasar
refractive scintillation as an useful, complementary probe of such
ionized, intergalactic gas. In particular, an application to the case
of the intracluster medium is presented.  We show that clusters located at $z
\approx 0.02$ should produce a source {\em rms} intensity fluctuation at
50-100 GHz of several tens of percent and on time scales ranging
from days to months, depending on the projected location of the source
on the foreground cluster. However, in order to produce such a signal,
the source needs to be very compact.  This effect, if observed, can
be used as an independent test of the baryonic mass fraction in clusters.

\end{abstract}

\section{Introduction}

Since the discovery of quasars, absorption line studies have been proposed     
to probe the prevailing conditions in the  intergalactic medium (IGM).
Initially (Gunn \& Peterson 1965, Sargent 1980, Blades, Turnshek \& Norman 1988),
most interest was focused on the redshifted Ly$\alpha$ resonance line
of neutral hydrogen arising in the so-called Ly$\alpha$ forest. More recently,
the technique has been extended to search for absorption lines associated with
ionized heavy elements (Cowie \etal 1995; Tytler \etal 1995; Lu \etal 1998;
Cowie \& Songaila 1998; Ellison \etal 1999, 2000) and it will likely be possible
to use it to investigate the existence of an analogous X-ray forest (Hellsten
\etal 1998, Perna \& Loeb 1998).

These experiments are sensitive to different states and physical processes
of the gas: Ly$\alpha$ lines can give information on the neutral component
of the IGM, whereas the other  two applications can be used to probe heavy
elements and the processes responsible for their ionization, likely
photo- and collisional ionization.
However, most of the baryonic mass in the universe is in the form of ionized
hydrogen: for example in Ly$\alpha$ clouds at $z \approx 3$ the neutral
fraction is typically only $10^{-5}$. At lower redshift, at least 40\% of the baryons
in the universe reside in a warm/hot component of the IGM
(Cen \& Ostriker 1999, Dav\'e \etal 2000).
Thus, it is certainly useful to think of independent ways to detect it.

As light waves traverse an ionized distribution of gas, they suffer a
phase change caused by variations in the average refractive index
related to random irregularities in the medium. This process is known
as scintillation, as, among other effects, it produces a wave
amplitude fluctuation.  Scintillation has been widely used at radio
frequencies in the study of the interstellar medium of the Milky Way
(MW) and a large literature is available (for a general introduction
we defer the reader to the excellent reviews by Rickett 1977 and 1990
and Goodman \& Narayan 1985). These experiments have brought a
tremendous advance towards the understanding of the spatial
distribution and small scale structure of the ionized component of the
ISM (see for example Armstrong, Rickett \& Spangler 1995 and
references therein). For example, local observations have established
that the power spectrum of density irregularities has a power-law
dependence on the wavenumber $\Phi_N(q) \propto q^{-\beta}$, with
$\beta =11/3$.  Intriguingly, this spectrum resembles the one expected
in the inertial range for a fully developed Kolmogorov turbulence.
However, recent work (Lambert \& Rickett 2000) suggests that at large
scale it might become steeper ($\beta \approx 4$), approaching the
spectrum produced by a random distribution of clumps or by a a complex
pattern of overlapping shock waves. A point worth stressing is that
there is no requirement that density irregularities be produced by
turbulence, although this is probably the case in the interstellar
medium of the MW. Indeed, any physical effect (\eg shocks,
photoionization, gravity, instabilities) producing a spatially
inhomogeneous ionized medium would also be necessarily responsible for
scintillation effects.

Hence, it seems that, if the sophisticated experimental
and theoretical tools developed for interstellar applications could be successfully
imported into IGM studies, scintillation could complement very interestingly the
ongoing absorption line observations.

It has to be pointed out that, being  scintillation insensitive to the gas
temperature as long as it is ionized, in principle it can be used to trace both
the  cool ($T=2-4\times 10^4$~K) gas residing either in the (mildly overdense)
Ly$\alpha$ clouds or in the halos of low mass ($10^8-10^9 M_\odot$) systems,
and the warm/hot gas ($T \ga 10^5$~K) originating from the virialization
of groups/clusters or shocked by cosmic explosions.

In spite of its excellent credit history and of the number of important
cosmological problems that it could potentially address,
scintillation due to intergalactic gas has received much  little attention. Application have
been made in the past to flickering of extragalactic sources (Rickett, Coles,
Bourgois 1984, Blandford, Narayan \& Romani 1986, Shishov 1989, Spangler \etal 1989,
Walker 1998), extreme scattering events (Fiedler \etal 1994) or, more recently,
to gamma-ray bursts (Goodman 1997; Frail \etal 1997; Waxman, Kulkarni \& Frail 1998).
However, virtually all such studies have been tried to infer properties of the
Galactic interstellar medium or of the source.

Here we take a different perspective and our aim is to show that
scintillation can provide fresh insights into the properties of
intergalactic medium by long-term radio monitoring of distant quasars;
to our knowledge, only a paper (Yoshioka 1970) has suggested such
possibility to date, although it has been hinted by a few  other
papers (Dennison \& Condon 1981; Goodman 1997, Cordes \& Lazio 2000). We show how the
scintillation signal produced by an intergalactic distribution of
ionized gas can be comparable to the standard interstellar one and,
more importantly, easily disentangled from that.  In this 
application we focus on the intracluster/intragroup gas.

\section{Scintillation: Basic Relations}

We consider scintillation of quasar light by a foreground cluster. Observations of
the $X$-ray surface brightness of galaxy clusters are commonly fitted
by the isothermal $\beta$ model, in which the gas temperature is
constant and the radial profile of its mass density is given by
\be
\rho_{gas}(r,T)=\rho_0(T)\left[1 + \left(\frac{r}{r_c}\right)^2\right]^{-3\beta/2},
\label{rhog}
\ee
where $r_c$ is the core radius. Typical values for the parameters of this
fit are clustered around $\beta\approx 2/3$ and $r_c\approx 0.25$ Mpc
(Jones \& Forman 1984), and we adopt these values in our calculations.
The central density is found by matching at large radii ($r\gg r_c$) the
density profile to that of an isothermal sphere. This yields
\be
\rho_0(T)=\frac{f_b kT}{2\pi\mu m_p Gr_c^2}\;,
\label{eq:rho0}
\ee
where $f_b=\Omega_b/\Omega$\footnote{We adopt an $\Omega=1$, $h=0.5$
cosmology throughout the discussion.} is the mass fraction
of the gas and $T$ its temperature, $m_p$ is the proton mass, and $\mu\approx 0.6$
is the mean atomic weight. We assign a cutoff radius $r_{\rm cut}$
so that the overdensity at the cluster boundary is about 10 times
the critical density.

Whereas it seems plausible that gas turbulent motions exist
in clusters, the turbulent spectrum has not been determined
so far. In the Galaxy, the power spectrum of the density fluctuations
is well described by the Kolmogorov  spectrum,
\be
\Phi_N(q)=C_N^2 q^{-\alpha -2}\;,
\label{eq:pn}
\ee
where $\alpha=5/3$. In our calculations, we assume this spectral form for the
gas density fluctuations over a range of scales from $l_0=10^{12}$ cm to
$l_1=10^{18}$ cm. This range is the typical one for the scales of
"refractive" inhomogeneities in the interstellar medium (Lambert \& Rickett 2000)
and, lacking better information, we assume it for the clusters as well.
The integral of $C_N^2$ along the line of sight is called the
scattering measure SM$\equiv\int_0^L dx C^2_N(x)$, where $L$ is the
distance to the source. In the problem that we are considering, it
can be written as
\be
{\rm SM}=2C_{N,0}^2\int_0^{x_{\rm max}}dx \frac{1}{\left[1+
\left(\frac{b^2+x^2}{r_c^2}\right)\right]^2}\;,
\label{eq:SM}
\ee
having chosen a system of coordinates where $x=0$ corresponds to the center
of the cluster.
Here, $b$ is the impact parameter (also measured from the cluster center),
$x_{\rm max}=\sqrt{r^2_{\rm cut}-b^2}$, and $C^2_{N,0}$ is the value of
$C^2_N$ at the center of the cluster. It can be written in terms of the
nominal Galactic value as $C^2_{N,0}=C^2_{N,{\rm Gal}}
[n_e(0)/0.02 \;{\rm cm}^{-3}]^2$, where $n_e(0)=\rho_0(T)/(\mu m_p)$, and
$C^2_{N,{\rm Gal}}=10^{-3.5}{\rm m}^{-20/3}$ (e.g. Goodman 1997).

The physically important quantity for scintillation is the longitudinal gradient
of the phase structure function
\be
D_\phi(s,z;\lambda)= 8\pi^2 r^2_e\lambda^2\int_{0}^{\infty}
\Phi_{N}(\kappa,z; q_z=0)[1- J_0(\kappa s)]
\kappa d\kappa\;,
\label{eq:Dphi}
\ee
where $\kappa=\sqrt{q_x^2+q_y^2}$, and $z$ is the direction of propagation
along the line of sight.
Here $c$ is the speed of light, $r_e$ the classic electron radius, and $\lambda$
the observation wavelength. For a plane wave incident on a scattering medium,
the line of sight integral of $D_\phi(s, z;\lambda)$ gives the structure
function of the geometric optics phase, or wave-structure function $D_s(s)$.
The field coherence length at the observer, $s_d$, is defined by $D_s(s_d)=1$.
The characteristic scattering angle is then defined by (see e.g. Goodman 1997) 
\be
\theta_d=\frac{\lambda}{2\pi s_d}=
2.93\;\nu_{10}^{-11/5}\;{\rm SM}_{-3.5}^{3/5}\;\mu \;{\rm a.s.}\;,
\label{eq:tetad}
\ee
where $\nu_{10}\equiv\nu/(10\;{\rm GHz})$ and 
SM$_{-3.5}\equiv {\rm SM}/(10^{-3.5}{\rm m}^{-20/3}
\;{\rm kpc})$.

For a pointlike source, $\theta_d$ can represent either a displacement in the
apparent position of the source, or a broadening of its angular size.
This depends on the magnitude of $\theta_d$ relative to the Fresnel angle
\be
\theta_F\equiv\left(\frac{\lambda}{2\pi d_{\rm scr}}\right)^{1/2}
\approx 2.57\; \nu_{10}^{-1/2}\;d_{\rm scr,kpc}^{-1/2}\;\mu \;{\rm a.s.}\;,
\label{eq:tetaf}
\ee
where $d_{\rm scr,kpc}$ is the distance to the screen in kpc.
If $\theta_d~<~\theta_F$, the scatterers can displace the image but not divide
it. This is the weak scattering regime. On the other hand, if $\theta_d >  \theta_F$,
then multiple images of the source are formed, and in this case $\theta_d$
can be interpreted as the effective size of the scatter-broadened source image.
This is the strong scattering regime.

Flux scintillation can occur in two forms, diffractive and refractive.
Diffractive scintillation is a physical-optics interference effect, and
can only occur if the scattering is strong, when multiple subimages of the
source are formed.  The ray path from source to observer is different for
each subimage, and if the receiving telescope is too small to separate these
subimages, then the observed flux is given by the coherent sum of the
contributions from all paths. Each subimage is analogous to a slit, and
on an imaginary plane perpendicular to the line of sight and passing through
the observer, the flux forms a speckle pattern. The characteristic size of a speckle
is $\lambda/(2\pi\theta_d)=s_d$. In order for the speckle pattern not to  be
smoothed out by source size effects, it is necessary that
the intrinsic angular size of the source $\theta_s$ be
$\theta_s< s_d/d_{\rm scr}$. When scintillation
is due to an extragalactic screen, this condition requires
a size which is much smaller than the emitting region of a quasar.
Therefore, diffractive scintillation is quenched, and
only refractive scintillation becomes possible in our context
(see also discussion below).

Refractive scintillation is a geometric-optics effect consisting in the
random magnification of individual subimages.   For an extended source it is
negligible until the scattering becomes very strong, and $\theta_d>\theta_s$.
The effective image of the source is given by the convolution of its
intrinsic surface brightness distribution with the scatter-broadened image
of a point source. This can be written as (Goodman 1997)
\be
\theta_{\rm eff}\equiv[\theta_s^2+(0.71\theta_d)^2+(0.85\theta_F)^2]^{1/2}\;.
\label{eq:tetaeff}
\ee
The effective image, being an incoherent source, can only be lensed
by fluctuations which are at least as large as its projection
on the scattering screen; this implies fluctuations of size
$s \simgt d_{\rm scr}\theta_{\rm eff}\equiv s_r$.
Inhomogeneities much larger than this scale are also ineffective
because they result in a weak scattering strength, \ie  low values of
$\theta_d/\theta_F$. Refractive scintillation
is therefore dominated by fluctuations on the order of $s_r$. This sets the typical timescale
of the flux modulation to
\be
t_{\rm r}=\frac{s_r}{v_\perp}\;.
\label{eq:tref}
\ee
In the case we are considering, this relative motion is dominated by the 
velocity of the inhomogeneities themselves. A reasonable lower limit for 
this value is given by the sound speed in the intracluster medium. Here we 
assume a cluster temperature $T=8$~keV, as appropriate for the Coma cluster 
(Mohr, Mathiesen \& Evrard 1999); 
the corresponding sound speed is roughly 1100 km~s$^{-1}$. Therefore we use 
$v_\perp \sim 10^3$~km~s$^{-1}$, which is also in agreement with the value
assumed for the turbulent velocity of the gas in  hot clusters by
Sarazin (1989). Note that the value of $v_\perp$ affects the 
refractive time scale linearly according to the above equation.

The strength of refractive scintillation is expressed in terms of the
root-mean-square flux variation relative to the mean, $m_R\equiv
\sqrt{C(0)}$, where $C({\bf s})\equiv \left<\delta I({\bf x}) \delta
I({\bf x}+{\bf s})\right>\;-\;1$ (with $\left< I \right>=1$) is the
intensity covariance; $m_r$ is referred to as the {\it modulation
index}.  In the situation that we are considering, where the 
scattering occurs in a cluster at cosmological distance
(but still much smaller than the distance to the source), we can adopt the
thin screen approximation, in which one assumes that the scattering 
is concentrated in a narrow layer at distance $d_{\rm scr}$. 
The modulation index can then be well approximated by (Goodman 1997)

\begin{equation}
m_{R} = 0.114 \nu_{10}^{-2} {\rm SM}_{-3.5}^{1/2}\;d_{\rm scr,kpc}^{-1/6}\;
\left(\frac{\theta_{\rm eff}}{10 \mu {\rm a.s.}}
\right)^{-7/6}\;,
\label{eq:mr}
\end{equation}

\section{Remarks on the Source Size}

An important assumption of our study concerns the existence of sources that
are compact enough to produce the refractive scintillation pattern that
we are predicting. This translates in the condition $\theta_s \simlt \theta_d$.
A discussion of this assumption seems worthwhile.

The intrinsic size of a source emitting {\it incoherent} synchrotron
radiation can be obtained from its brightness temperature $T_b$. Radio
observations typically test the Rayleigh part of the black-body curve. Hence
we have
\begin{equation}
B_\nu(T_b) = {2 k \nu^2 T_b\over c^2},
\end{equation}
which defines $T_b$. Equating $B_\nu$ to the measured flux of the source,
$S_\nu$ divided by the solid angle subtended by the source, $\pi \theta_s^2$
one obtains\footnote{We use the notation $\nu_X = (\nu/X)$ for the measure units}
\begin{equation}
\theta_s = 2.1 \times 10^7 \left( {S_{\nu,mJy}\over T_b \nu_{GHz}^2}\right)^{1/2} \mu
{\rm as}.
\end{equation}
The expression for $\theta_d$ is given by Equation~(\ref{eq:tetad}).
Using the typical values SM=SM$_{-3.5}$  and $S_{\nu,mJy}=100$, we get
\begin{equation}
{\theta_d\over \theta_s} = 2.2 \times 10^{-6} {T_b^{1/2} \over \nu_{GHz}^{6/5}}.
\end{equation}
This implies that for the above ratio to be larger than, say, three, one needs $T_b$ to
be in excess of few $\times 10^{16}$~K. Upper limits on $T_b$ are given by the
condition on inverse Compton cooling (Kellerman \& Pauliny-Toth 1969) which
imply that the brightness temperature for incoherent synchrotron emission
has to be lower than $\approx 10^{12}$~K. Above that value, sources would undergo
catastrophic cooling; this condition therefore represents a (wavelength independent)
theoretical limit on $T_b$.

However, many sources are observed for which a $T_b$ higher than this
value has been estimated (Romero 1994, Quirrenbach \etal 1992,
Ter\"asranta \& Valtoja 1994). Observationally, it
is well established that these sources, known as blazars, commonly
show variability on time scales of hours (the so-called IntraDay
Variability [IDV], for a review see Wagner \& Witzel 1995). If this
variability is internal, then the upper limit on the size is $ c \tau
\approx c \times {\rm (hours)} \approx 10^{14}$~cm. At cosmological
distances of about 1 Gpc, this corresponds to $6.7 \times 10^{-3}
~\mu.$a.s. Consequently, inferred values of the brightness
temperature are in the range $T_b \sim 10^{18-21}$~K.

One possible explanation (Rees 1966; Woltjer 1966) is that
these sources are relativistically beamed by a Doppler factor ${\cal D} = [\Gamma
(1-\beta \cos\theta)^{-1}]$. As $T_b$ scales as ${\cal D}^3$, one requires ${\cal D}
\approx 10^3$ to account for the highest $T_b$ sources. These Doppler factors are
however too high with respect to those inferred from superluminal motions ($< 20$)
and for  $\gamma$-ray fluxes to be compatible with $\gamma\gamma$ pair absorption,
also implying ${\cal D} < 20$.

Therefore, as the required values of $T_b$ (even allowing for a very large
Doppler factor) implied by this compactness are implausibly large, one
has to admit that, at least for this class of sources, the observed
radiation is {\it coherent}, so that the above discussed Compton limit
does not apply.  Alternatively, one can note that IDV can have an
external origin, for example microlensing or scintillation by the ISM
(time scales are too short for the IGM to contribute appreciably). One
of the most stringent arguments in support of this interpretation is
that low-frequency variable sources are almost always located behind large
scale galactic structures (Shapirovskaya 1978). Hence, scintillation
implies that at these frequencies the source is compact enough to
produce refractive scintillation on such short timescales.
The extreme brightness temperatures of $10^{21}$K
that would be deduced by an intrinsic interpretation
of  the fastest IDV were rejected in the papers
announcing the observations (\eg Kedziora-Chudczer \etal 1997, 
and Dennett-Thorpe \& de Bruyn 2000).
These authors favor an ISS model; a view also
supported by Rickett \etal (1995) for a more conventional
IDV source.
At higher frequencies, if the necessary compactness is
maintained is not known. In any case, at present there are no
theoretical or observational arguments that can rule out this
possibility, and IDV blazars appear to represent the most suitable
background candidates for our study.

\section{Results}

We start by analyzing the behavior of the refractive time scale (Fig. 1),
$t_{\rm r}$, as a function of the impact parameter $b$ (in units of $r_c$) at which
the line of sight to a quasar intersects the cluster hot gas
distribution. In the cases we discuss, the quasar (source) is
assumed to be located at $z_s=1.0$ whereas the cluster (screen)
is taken to be at $z_c=0.02$.   
We also consider two different values for $f_b$ (0.04, 0.2) that
should encompass a reasonable variation range for this parameter;
for the observation frequency we consider the cases of 100~GHz and 50~GHz.
The time scale of refractive scintillation 
ranges from about a day to several months, depending on the position within the
cluster and the frequency of observation. Here we have considered the case
of a point source, for which $\theta_s\ll \theta_d, \theta_F$. The refractive 
time scale is then determined by the largest between $\theta_d$ and $\theta_F$.
In the inner part of the cluster, $\theta_d\ga \theta_F$, and $t_{\rm ref}\propto
\theta_d\propto \nu^{-11/5} {\rm SM}^{3/5}$. The time scales in this region
are very sensitive to the value of SM, and therefore of $f_b$. In the
outer part, on the other hand, $\theta_F$ is larger than $\theta_d$, and
the refractive time scale becomes $t_{\rm ref}\propto \theta_F\propto \nu^{-1/2}
d_{\rm scr}^{1/2}$, which is independent of SM, but is essentially
determined by the frequency of observation and the distance to the screen.

Changing $z_{\rm source}$ in the range $z\sim1-5$ does not appreciably
change the results.  The reason is that the distance to the source
only enters the definition of the source intrinsic angular size, which
we assumed here to be negligible with respect to $\theta_d$ in the
expression of $\theta_{eff}$.  The results are more sensitive to
variations in $z_{\rm scr}$, and the time scale roughly increases
as $(1+z_{\rm scr})$.  As the observation frequency decreases, the
refractive time scales increases.  Hence, it appears that frequencies
in the range considered here provide the best detection prospectives
for extragalactic scintillation. 

In Fig. 2, we show the refractive modulation index $m_r$ calculated
for the same observation frequencies, $100$~GHz and $50$~GHz.  The
modulation index is generally on the order of several tens of
percent, and it varies significantly with both frequency and
baryonic fraction $f_b$. In the inner region of
the cluster, scintillation is in the strong regime; here
$\theta_{\rm eff} \simeq \theta_d$, and Equation~(\ref{eq:mr}) is
reduced to $m_R\simeq 0.7\nu_{10}^{17/30} {\rm SM}_{-3.5}^{-1/5}
d_{\rm scr,kpc}^{-1/6}$.  The intensity of the modulation increases
with frequency and decreases with SM (and therefore with $f_b$).  This
behaviour is inverted in the outer parts of the cluster, where the
regime becomes weak. Here $\theta_{\rm eff} \simeq \theta_F$, and
Equation~(\ref{eq:mr}) is roughly given by $m_R\simeq
0.67\nu_{10}^{-17/12} {\rm SM}_{-3.5}^{1/2} d_{\rm
scr,kpc}^{5/12}$. Note that, even though the regime here is weak and
the scattering measure is smaller than the galactic value, there can still be a
relatively strong signal due to the fact that the screen is at a
cosmological distance.

As in Figure 1, the results in Figure 2 have been computed for the
case of a very compact source, and this is a crucial assumption in
order to have a significant signal. In fact, as the size of the source
increases and becomes larger than $\theta_d$, $\theta_F$, the
amplitude of the fluctuations is reduced. In the limit where
$\theta_s\gg \theta_d, \theta_F$, one has $\theta_{\rm eff}\simeq
\theta_s$, and, using Equation~(\ref{eq:mr}), one can see that the
modulation index is roughly given by $m_R^{\rm ext}\simeq m_R^{\rm
point} (\theta_{d}/\theta_s)^{7/6}$ in the strong scattering regime,
and by $m_R^{\rm ext}\simeq m_R^{\rm point}
(\theta_{F}/\theta_s)^{7/6}$ in the weak regime. The suppression in
the modulation of the signal as the size of the source is increased is
explicitly shown in Figure 3. For a given source size $l_s$, the
amount of suppression is different in the various parts of the
cluster. In the innermost regions, where $\theta_d$ is larger, source
size effects become noticeable for $l_s\ga 10^{16}$ cm. In the outer
parts, where the ratio $\theta_d/\theta_s$ is smaller, the
suppression becomes evident for smaller sizes and, for a source of
$\sim10^{16}$ cm at $z_s=1$, the modulation of the signal is reduced
to only a few percent.  Furthermore, note that, while reducing the
modulation, a large source also makes the refractive time scale
longer, of a factor on the order of $\theta_s/\theta_d$ in the strong
regime, and of $\theta_s/\theta_F$ in the weak case.

\section{Discussion}

We have presented an extension of standard scintillation experiments
to probe the intergalactic medium; in particular, we have considered
the case of intracluster gas. We have found that clusters located at
redshift around a tenth should produce an rms intensity fluctuation at
50-100 GHz of several tens of percent on time scales ranging from
about a day to several months, depending on the projected location of the
source with respect to the cluster radius.  This effect, if observed,
can yield valuable information on the baryonic fraction
($\Omega_b/\Omega$) and the spectrum of inhomogeneities of the
intracluster gas.

More generally, this technique could be particularly important to
study the warm/hot gas predicted to act as a reservoir of the majority
of baryons at low redshift. In fact, we have mentioned that
scintillation is not sensitive to temperature, as long as the gas is
ionized. However, only the largest objects (clusters and groups of
galaxies) can reach a sufficiently high scattering measure to produce 
a significant fluctuation level. Thus, it appears that scintillation
is a perfect tool to investigate hot gas, presumably in large
virialized objects. Constraints on the baryonic mass fraction in
clusters that can complement those obtained from emission/absorption
line measurements and from the Sunyaev-Zeldovich effect should be
obtainable if scintillation will be observed.  Scintillation is
particularly effective to probe gas at very large radii, which is
difficult to probe with $X$-ray instruments.

The proposed method can be also very important in assessing the level of turbulence
in the intracluster medium, if the data will come out to be consistent with our
predictions based on the assumption that a density fluctuation spectrum is close
to the one expected for Kolmogorov turbulence. Turbulence has been often invoked
in the cluster environment to provide efficient mechanism for heavy element mixing;
it can be generated both by stirring of the gas by galaxy motions, shocks
(both due to virialization and by galactic outflows), mergings and also by
Kelvin-Helmoltz instabilities occurring at the boundaries of galaxies as they move
through the intracluster medium (Goldman \& Raphaeli 1991, de Young 1992,
Westbury \& Henriksen 1992, Goldshmidt \& Raphaeli 1993, Sanchez-Salcedo,
Brandenburg \& Shukurov 1998, Mori \& Burkert 2000). However, this phenomenon is
still awaiting firm, direct detections.
We have shown that intergalactic scintillation might well provide one.

Observations on long time scales obviously require some care
as they can be affected by several spurious effects.  Flux modulations
due to scintillation should however be easily isolated from flux variations
due to other causes.  Amplifications due to microlensing are
achromatic, and therefore distinguishable from the frequency dependent
scintillation, whereas intrinsic variations are not expected to be of
periodic nature and to have the same frequency dependence as
scintillation.  Multiwavelength observations, as already available for
several objects (e.g. Bloom et al. 1999), would therefore be most
useful to such purpose.  If more than one source were found behind a
cluster, then the dependence of the signal on the impact parameter would
also be very distinctive of scintillation.  

A contribution to refractive scintillation by the galactic ISM is also possible.
However, for most of the cluster (where the regime is weak and
$m_R\propto {\rm SM}^{1/2} d_{\rm scr}^{5/12}$), the contribution  
to the modulation index by the IGM should largely dominate over that from
the ISM. For a cluster at $z\sim 0.02$, we have $d_{\rm IGM}\sim 10^5 
d_{\rm Gal}$, and therefore the contribution to $m_R$ from the IGM will dominate over
that from the ISM for $SM_{\rm IGM}\ga 10^{-5} SM_{\rm Gal}$.   
At the same time, the refractive time scale at 100 GHz, obtained assuming a screen
with SM=SM$_{Gal}$ located at 1 kpc from the observer, is on the order
of an hour, thus much shorter than the one due to the hot cluster gas.

In principle, one can also think of applications to the ``true'' IGM, 
\ie the diffuse intergalactic gas observed via  Ly$\alpha$ absoption lines. 
For an Einstein-de Sitter universe with $\Omega_bh^2 =0.019$, the
IGM scattering measure due to a uniformly distributed IGM up to a 
redshift $z$ is $2.8 \times 10^{-5} (1+z)^{9/2} SM_{\rm Gal}$, implying
that a contribution can occur only for sources located at very high $z$.
In addition, the typical Doppler parameters observed in the Ly$\alpha$ 
forest are at most a few tens of km~s$^{-1}$, thus making $t_r$ much longer. 
We conclude that a scintillation component arising from the true IGM would be 
hardly detectable.

An important assumption in our discussion regards the compactness of
the source. We have discussed the decrease of $m_r$ when the source
size is increased and we have shown (Fig. 3) that  compact sources 
are needed in order to have a significant modulation of the signal. 

Using eq. (12), it is easy to show that the relation linking the source
size $l_s$ with its brightness temperature is 
\be 
T_b = 3.6 \times
10^{15} S_{\nu,mJy} \nu_{GHz}^{-2}\ell_{s,16}^{-2} {\rm~ K}\;, 
\label{bright}
\ee
which gives $T_b=3.6 \times 10^{13}$~K for $S_{\nu,mJy}=100$,
$\nu_{GHz}=100$, and $l_s=10^{16}$ cm.  
For these values, the modulation index is $\sim 8\%$ at the cluster
center, and $\sim 2\%$ when SM=SM$_{\rm gal}$.
What we now need to estimate
is the surface density in the sky of the IDV sources which have the required
brightness temperature.
IDV in radio-loud sources has only been seen in
objects with flat spectra (Flat Spectrum Radio Quasars, FSRQ), \ie
$S_\nu \propto \nu^{\alpha}, \alpha > 0.5$; among these sources about
25\% show IDV (Ghisellini, private communication).  For the
differential number counts of FSRQ we use the results of Padovani \&
Urry (1992) (see their Fig. 5); at $S_{\nu,mJy}=100$ they find 108
such sources (or 27 IDV sources) per steradian. Thus, the probability
of finding one such object behind the central $\approx 2$ Mpc region of
our reference cluster at $z=0.02$ is about 3\%, i.e. a small but
nonnegligible occurrence chance.

\section*{Acknowledgments}
This work was completed as one of us (AF) was a Visiting Professor at
the Center for Computational Physics, Tsukuba University, whose
support is gratefully acknowledged. We thank A. Celotti,
G. Ghisellini, H. Susa and M. Umemura for useful discussions.  Most of
all, we are indebted to the referee, B. J. Rickett, whose careful
reading and insightful comments greatly improved our manuscript.

\label{lastpage}

\newpage

\begin{figure*}
\psfig{figure=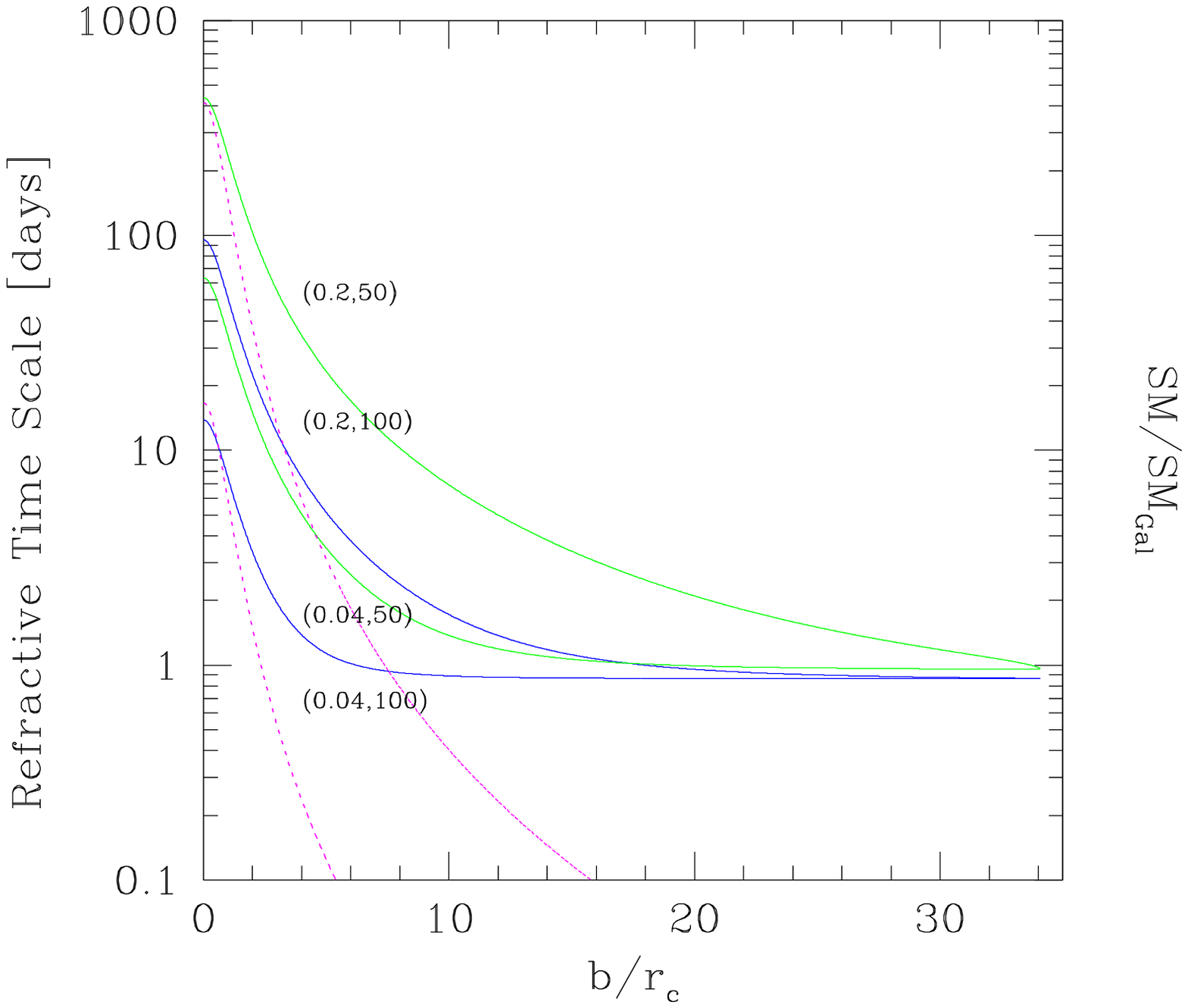}
\caption{\label{fig1}\footnotesize{Refractive time scale $t_{\rm r}$ (solid curves)
and scattering measure $SM$ (dotted) for the observed frequencies $\nu=50, 100$~GHz,
as a function of the line of sight impact parameter
$b$ in units of $r_c$. Both curves are shown for two values of the baryon fraction.
The label on top of each curve indicates ($f_b$, $\nu_{\rm GHz}$).
}}
\end{figure*}

\begin{figure*}
\psfig{figure=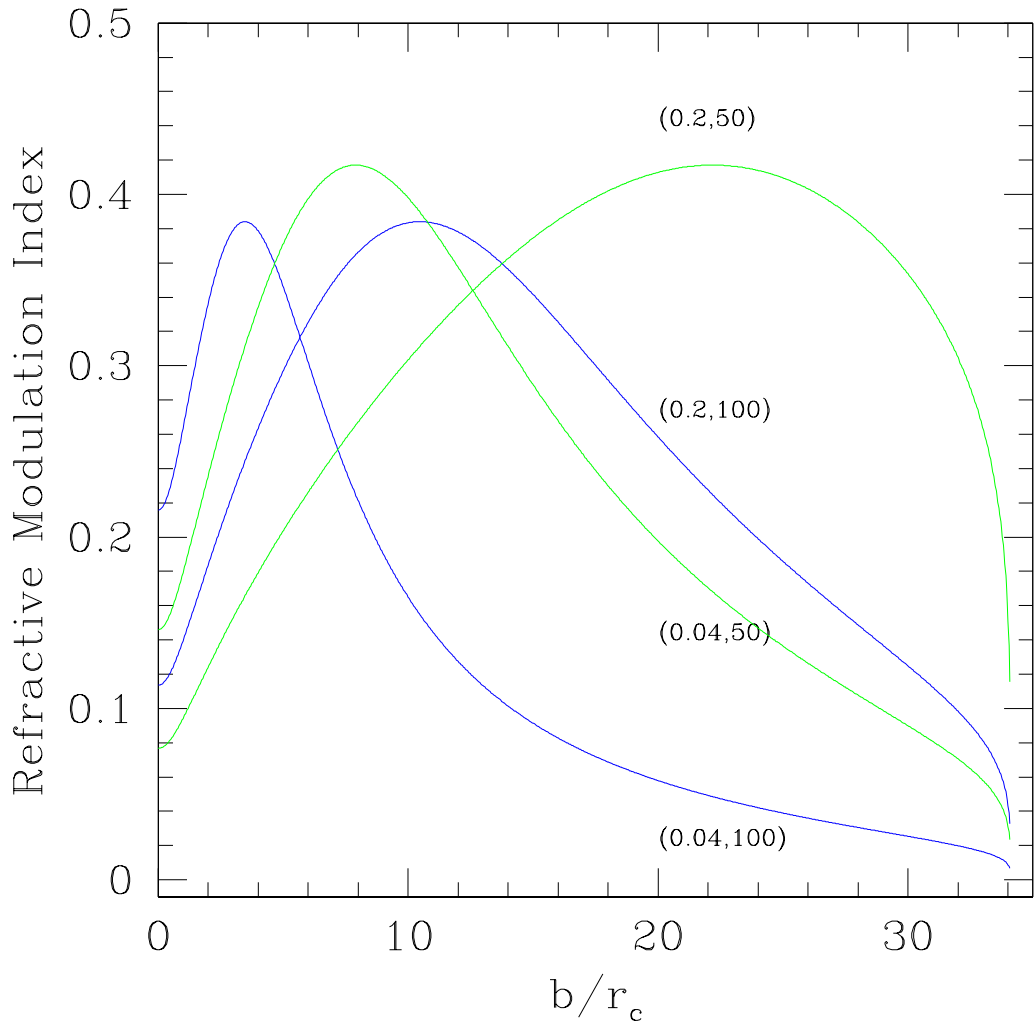}
\caption{\label{fig2}\footnotesize{Refractive modulation index $m_r$
as a function of the line of sight impact parameter
$b$ in units of $r_c$. Curves are shown for two values of the baryon fraction
and two frequencies, which are indicated near each curve using the same
notation as in Figure 1.}}
\end{figure*}

\begin{figure*}
\psfig{figure=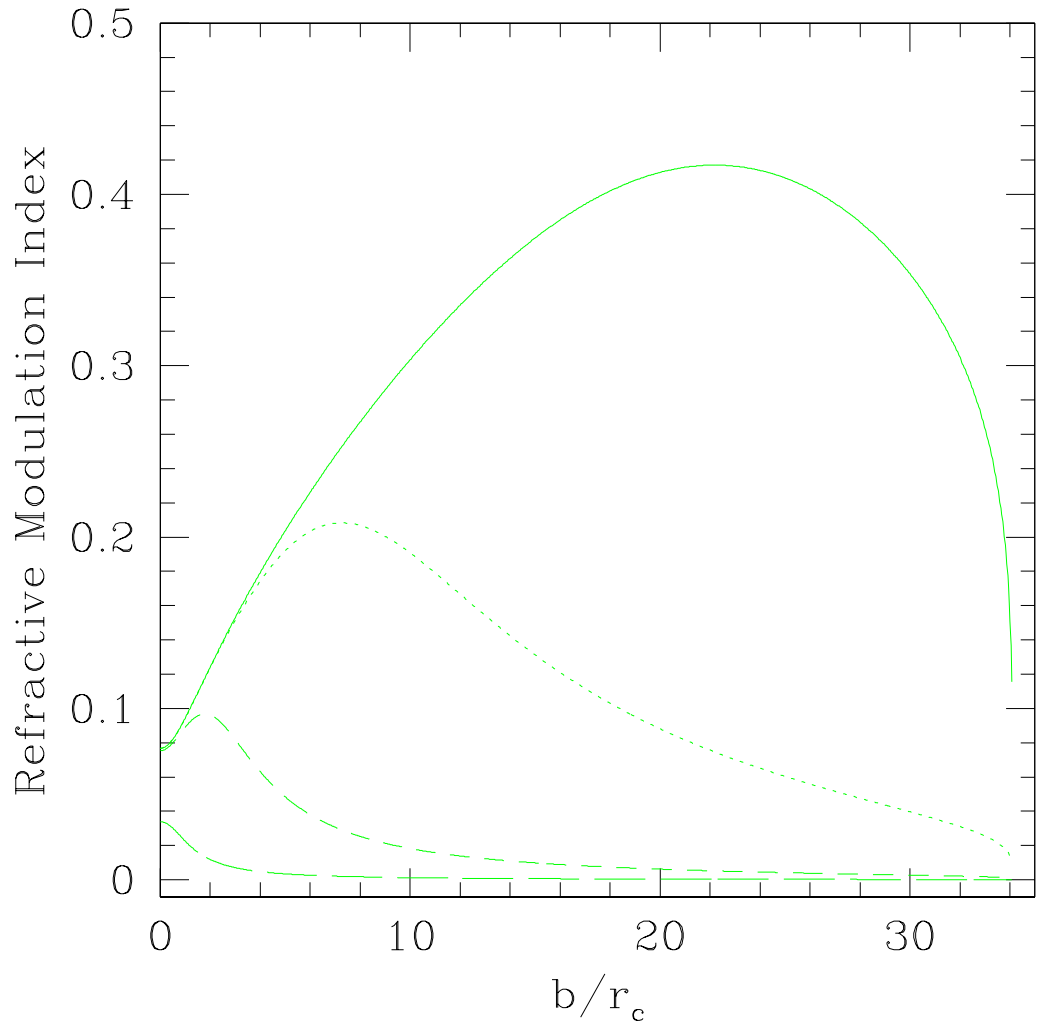}
\caption{Modulation index for $\nu=50$ GHz, $f_b=0.2$ and for
various source sizes, $l_s$ =  $10^{14}$ cm (solid line), $10^{15}$ cm
(dotted line), $10^{16}$ cm (dashed line), and $10^{17}$ cm
(long-dashed line). The position of the source is at $z_s=1$.}  

\end{figure*}


\end{document}